\documentclass[conference]{IEEEtran}
\IEEEoverridecommandlockouts
\usepackage[moderate,tracking=normal]{savetrees}
\usepackage[backend=bibtex,giveninits=true,hyperref=true,maxnames=3,minnames=1,natbib=true,style=savetrees,sorting=none]{biblatex}
\AtBeginBibliography{\small}
\usepackage{amsmath,amssymb,amsfonts}
\usepackage{mathtools}
\usepackage{graphicx}
\usepackage{textcomp}
\usepackage{xcolor}
\usepackage{multirow}
\usepackage{subcaption}
\usepackage[T1]{fontenc}
\usepackage[labelfont=bf,font=small]{caption}
\usepackage[nolist,printonlyused]{acronym}
\usepackage{comment}

\usepackage[compact]{titlesec}

\newcommand{\mx}[1]{\mathbf{#1}}

\newcommand{\xhdr}[1]{\vspace{1.7mm}\noindent{{\bf #1.}}}

\newtheorem{lemma}{Lemma}

\def\BibTeX{
{\rm B\kern-.05em{\sc i\kern-.025em b}\kern-.08em
T\kern-.1667em\lower.7ex\hbox{E}\kern-.125emX}
}
\usepackage[compact]{titlesec}
\usepackage{cuted} 

\begin{acronym}
\acro{NF}{near-field}
    \acro{PR}{partial relaxation}
\acro{PR-ML}{partial-relaxation maximum likelihood}
\acro{PR-CF}{partial-relaxation covariance fitting}
  \acro{ULA}{uniform linear array}
  \acro{CRB}{Cram\'er--Rao bound}
  \acro{CCRB}{constrained Cram\'er--Rao bound}
  \acro{ML}{maximum likelihood}
  \acro{SNR}{signal-to-noise ratio}
\acro{2D-MUSIC}{two-dimensional MUltiple SIgnal Classification}
  \acro{PR-CRB}{partial-relaxation Cram\'er--Rao bound}
  \acro{PR-CCRB}{partial-relaxation constrained Cram\'er--Rao bound}
\acro{FIM}{Fisher information matrix}
\acro{PRCRBp}[PR-CRB\ensuremath{_{\mathrm{p}}}]
{Partial-Relaxation Cram\'er--Rao Bound for Known Pilots}

\acro{PRCRBu}[PR-CRB\ensuremath{_{\mathrm{u}}}]
{Partial-Relaxation Cram\'er--Rao Bound for Unknown Symbols}
\acro{NMS}[NMS]{non-maximum suppression}

  \acro{2G}{Second Generation}
  \acro{3G}{3$^\text{rd}$~Generation}
  \acro{3GPP}{3$^\text{rd}$~Generation Partnership Project}
  \acro{4G}{4$^\text{th}$~Generation}
  \acro{5G}{5$^\text{th}$~Generation}
  \acro{AA}{Antenna Array}
  \acro{AC}{Admission Control}
  \acro{AD}{Attack-Decay}
  \acro{ADP}{Angle-Delay Profile}
  \acro{ADSL}{Asymmetric Digital Subscriber Line}
	\acro{AHW}{Alternate Hop-and-Wait}
  \acro{AMC}{Adaptive Modulation and Coding}
  \acro{AoA}{angle of arrival}
  \acro{AoD}{angle of departure}
	\acro{AP}{Access Point}
  \acro{APA}{Adaptive Power Allocation}
  \acro{AR}{autoregressive}
  \acro{ARMA}{Autoregressive Moving Average}
  \acro{ATES}{Adaptive Throughput-based Efficiency-Satisfaction Trade-Off}
  \acro{AWGN}{additive white Gaussian noise}
  \acro{BB}{Branch and Bound}
  \acro{BD}{Block Diagonalization}
  \acro{BER}{bit error rate}
  \acro{BF}{Best Fit}
  \acro{BLER}{BLock Error Rate}
  \acro{BPC}{Binary power control}
  \acro{BPSK}{Binary Phase-Shift Keying}
  \acro{BPA}{Best \ac{PDPR} Algorithm}
  \acro{BRA}{Balanced Random Allocation}
  \acro{BCRB}{Bayesian Cram\'{e}r-Rao Bound}
  \acro{BS}{base station}
  \acro{CAP}{Combinatorial Allocation Problem}
  \acro{CAPEX}{Capital Expenditure}
  \acro{CBF}{Coordinated Beamforming}
  \acro{CBR}{Constant Bit Rate}
  \acro{CBS}{Class Based Scheduling}
  \acro{CC}{Congestion Control}
  \acro{CDF}{Cumulative Distribution Function}
  \acro{CDMA}{Code-Division Multiple Access}
  \acro{CIR}{Channel Impulse Response}
  \acro{CL}{Closed Loop}
  \acro{CLPC}{Closed Loop Power Control}
  \acro{CN}{Core Network}
  \acro{CNR}{Channel-to-Noise Ratio}
  \acro{CPA}{Cellular Protection Algorithm}
  \acro{CPICH}{Common Pilot Channel}
  \acro{CoMP}{Coordinated Multi-Point}
  \acro{CQI}{Channel Quality Indicator}
  \acro{CRM}{Constrained Rate Maximization}
	\acro{CRN}{Cognitive Radio Network}
  \acro{CS}{Coordinated Scheduling}
  \acro{CSI}{channel state information}
  \acro{CSIR}{channel state information at the receiver}
  \acro{CSIT}{channel state information at the transmitter}
  \acro{CUE}{cellular user equipment}
  \acro{D2D}{device-to-device}
  \acro{DCA}{Dynamic Channel Allocation}
  \acro{DE}{Differential Evolution}
  \acro{DFT}{Discrete Fourier Transform}
  \acro{DIST}{Distance}
  \acro{DL}{downlink}
  \acro{DMA}{Double Moving Average}
	\acro{DMRS}{demodulation reference signal}
  \acro{D2DM}{D2D Mode}
  \acro{DMS}{D2D Mode Selection}
  \acro{DPC}{Dirty Paper Coding}
  \acro{DRA}{Dynamic Resource Assignment}
  \acro{DSA}{Dynamic Spectrum Access}
  \acro{DSM}{Delay-based Satisfaction Maximization}
  \acro{ECC}{Electronic Communications Committee}
  \acro{EFLC}{Error Feedback Based Load Control}
  \acro{EI}{Efficiency Indicator}
  \acro{eNB}{Evolved Node B}
  \acro{EPA}{Equal Power Allocation}
  \acro{EPC}{Evolved Packet Core}
  \acro{EPS}{Evolved Packet System}
  \acro{ESPRIT}{estimation of signal parameters via rotational invariance}
  \acro{E-UTRAN}{Evolved Universal Terrestrial Radio Access Network}
  \acro{ES}{Exhaustive Search}
  \acro{FDD}{frequency division duplexing}
  \acro{FDM}{Frequency Division Multiplexing}
  \acro{FER}{Frame Erasure Rate}
  \acro{FF}{Fast Fading}
  \acro{FI}{Fisher information}
  \acro{FIM}{Fisher information matrix}
  \acro{FSB}{Fixed Switched Beamforming}
  \acro{FST}{Fixed SNR Target}
  \acro{FTP}{File Transfer Protocol}
  \acro{GA}{Genetic Algorithm}
  \acro{GBR}{Guaranteed Bit Rate}
  \acro{GLR}{Gain to Leakage Ratio}
  \acro{GNN}{Graph Neural Network}
  \acro{GCN}{Graph Convolution Network}
  \acro{GOS}{Generated Orthogonal Sequence}
  \acro{GPL}{GNU General Public License}
  \acro{GRP}{Grouping}
  \acro{HARQ}{Hybrid Automatic Repeat Request}
  \acro{HMS}{Harmonic Mode Selection}
  \acro{HOL}{Head Of Line}
  \acro{HSDPA}{High-Speed Downlink Packet Access}
  \acro{HSPA}{High Speed Packet Access}
  \acro{HTTP}{HyperText Transfer Protocol}
  \acro{ICMP}{Internet Control Message Protocol}
  \acro{ICI}{Intercell Interference}
  \acro{ID}{Identification}
  \acro{IDFT}{Inverse Discrete Fourier Transform}
  \acro{ISAC}{integrated sensing and communication}
  \acro{IETF}{Internet Engineering Task Force}
  \acro{ILP}{Integer Linear Program}
  \acro{JRAPAP}{Joint RB Assignment and Power Allocation Problem}
  \acro{UID}{Unique Identification}
  \acro{IID}{Independent and Identically Distributed}
  \acro{IIR}{Infinite Impulse Response}
  \acro{ILP}{Integer Linear Problem}
  \acro{IMT}{International Mobile Telecommunications}
  \acro{INV}{Inverted Norm-based Grouping}
	\acro{IoT}{Internet of Things}
  \acro{IP}{Internet Protocol}
  \acro{IPv6}{Internet Protocol Version 6}
  \acro{ISD}{Inter-Site Distance}
  \acro{ISI}{Inter Symbol Interference}
  \acro{ITU}{International Telecommunication Union}
  \acro{JOAS}{Joint Opportunistic Assignment and Scheduling}
  \acro{JOS}{Joint Opportunistic Scheduling}
  \acro{JP}{Joint Processing}
	\acro{JS}{Jump-Stay}
  \acro{KKT}{Karush-Kuhn-Tucker}
  \acro{KPI}{key performance indicator}
  \acro{L3}{Layer-3}
  \acro{LAC}{Link Admission Control}
  \acro{LA}{Link Adaptation}
  \acro{LC}{Load Control}
  \acro{LoS}{line-of-sight}
  \acro{LOS}{Line of Sight}
  \acro{LP}{Linear Programming}
  \acro{LS}{least squares}
  \acro{LTE}{Long Term Evolution}
  \acro{LTE-A}{LTE-Advanced}
  \acro{LTE-Advanced}{Long Term Evolution Advanced}
  \acro{M2M}{Machine-to-Machine}
  \acro{MAC}{Medium Access Control}
  \acro{MAE}{Mean Absolute Error}
  \acro{MANET}{Mobile Ad hoc Network}
  \acro{MC}{Modular Clock}
  \acro{MCS}{Modulation and Coding Scheme}
  \acro{MDB}{Measured Delay Based}
  \acro{MDI}{Minimum D2D Interference}
  \acro{MF}{Matched Filter}
  \acro{MG}{Maximum Gain}
  \acro{MH}{Multi-Hop}
  \acro{MIMO}{multiple input multiple output}
  \acro{MINLP}{Mixed Integer Nonlinear Programming}
  \acro{MIP}{Mixed Integer Programming}
  \acro{MISO}{Multiple Input Single Output}
  \acro{ML}{machine learning}
  \acro{MLE}{maximum likelihood estimator}
  \acro{MLWDF}{Modified Largest Weighted Delay First}
  \acro{MME}{Mobility Management Entity}
  \acro{MMSE}{minimum mean squared error}
  \acro{MOS}{Mean Opinion Score}
  \acro{MPC}{Multi-path Component}
  \acro{MPF}{Multicarrier Proportional Fair}
  \acro{MRA}{Maximum Rate Allocation}
  \acro{MR}{Maximum Rate}
  \acro{MRC}{Maximum Ratio Combining}
  \acro{MRT}{Maximum Ratio Transmission}
  \acro{MRUS}{Maximum Rate with User Satisfaction}
  \acro{MS}{mobile station}
  \acro{MSE}{mean squared error}
  \acro{MSI}{Multi-Stream Interference}
  \acro{MTC}{Machine-Type Communication}
  \acro{MTSI}{Multimedia Telephony Services over IMS}
  \acro{MTSM}{Modified Throughput-based Satisfaction Maximization}
  \acro{MU-MIMO}{multiuser multiple input multiple output}
  \acro{MU}{multi-user}
  \acro{MUSIC}{MUltiple SIgnal Classification}
  \acro{NAS}{Non-Access Stratum}
  \acro{NB}{Node B}
  \acro{NE}{Nash equilibrium}
  \acro{NCL}{Neighbor Cell List}
  \acro{NLP}{Nonlinear Programming}
  \acro{NLOS}{Non-Line of Sight}
  \acro{NN}{Nearest Neighbor}
  \acro{NMSE}{normalized mean squared error}
  \acro{NORM}{Normalized Projection-based Grouping}
  \acro{NP}{Non-Polynomial Time}
  \acro{NR}{New Radio}
  \acro{NRT}{Non-Real Time}
  \acro{NSPS}{National Security and Public Safety Services}
  \acro{O2I}{Outdoor to Indoor}
  \acro{OFDMA}{orthogonal frequency division multiple access}
  \acro{OFDM}{orthogonal frequency division multiplexing}
  \acro{OFPC}{Open Loop with Fractional Path Loss Compensation}
	\acro{O2I}{Outdoor-to-Indoor}
  \acro{OL}{Open Loop}
  \acro{OLPC}{Open-Loop Power Control}
  \acro{OL-PC}{Open-Loop Power Control}
  \acro{OPEX}{Operational Expenditure}
  \acro{ORB}{Orthogonal Random Beamforming}
  \acro{JO-PF}{Joint Opportunistic Proportional Fair}
  \acro{OSI}{Open Systems Interconnection}
  \acro{PAIR}{D2D Pair Gain-based Grouping}
  \acro{PAPR}{Peak-to-Average Power Ratio}
  \acro{P2P}{Peer-to-Peer}
  \acro{PC}{Power Control}
  \acro{PCI}{Physical Cell ID}
  \acro{PDF}{Probability Density Function}
  \acro{PDPR}{pilot-to-data power ratio}
  \acro{PER}{Packet Error Rate}
  \acro{PF}{Proportional Fair}
  \acro{P-GW}{Packet Data Network Gateway}
  \acro{PL}{Pathloss}
  \acro{PPR}{pilot power ratio}
  \acro{PRB}{physical resource block}
  \acro{PROJ}{Projection-based Grouping}
  \acro{ProSe}{Proximity Services}
  \acro{PS}{Packet Scheduling}
  \acro{PSAM}{pilot symbol assisted modulation}
  \acro{PSK}{phase-shift keying}
  \acro{PSO}{Particle Swarm Optimization}
  \acro{PZF}{Projected Zero-Forcing}
  \acro{QAM}{Quadrature Amplitude Modulation}
  \acro{QoS}{Quality of Service}
  \acro{QPSK}{Quadri-Phase Shift Keying}
  \acro{RAISES}{Reallocation-based Assignment for Improved Spectral Efficiency and Satisfaction}
  \acro{RAN}{Radio Access Network}
  \acro{RA}{Resource Allocation}
  \acro{RAT}{Radio Access Technology}
  \acro{RATE}{Rate-based}
  \acro{RB}{resource block}
  \acro{RBG}{Resource Block Group}
  \acro{REF}{Reference Grouping}
  \acro{RF}{Radio Frequency}
  \acro{RLC}{Radio Link Control}
  \acro{RM}{Rate Maximization}
  \acro{RNC}{Radio Network Controller}
  \acro{RND}{Random Grouping}
  \acro{RRA}{Radio Resource Allocation}
  \acro{RRM}{Radio Resource Management}
  \acro{RSCP}{Received Signal Code Power}
  \acro{RSRP}{Reference Signal Receive Power}
  \acro{RSRQ}{Reference Signal Receive Quality}
  \acro{RR}{Round Robin}
  \acro{RRC}{Radio Resource Control}
  \acro{RSSI}{Received Signal Strength Indicator}
  \acro{RT}{Real Time}
  \acro{RU}{Resource Unit}
  \acro{RUNE}{RUdimentary Network Emulator}
  \acro{RV}{Random Variable}
  \acro{Rx}{receiver}
  \acro{SAC}{Session Admission Control}
  \acro{SCM}{Spatial Channel Model}
  \acro{SC-FDMA}{Single Carrier - Frequency Division Multiple Access}
  \acro{SD}{Soft Dropping}
  \acro{S-D}{Source-Destination}
  \acro{SDPC}{Soft Dropping Power Control}
  \acro{SDMA}{Space-Division Multiple Access}
  \acro{SER}{Symbol Error Rate}
  \acro{SES}{Simple Exponential Smoothing}
  \acro{S-GW}{Serving Gateway}
  \acro{SINR}{signal-to-interference-plus-noise ratio}
  \acro{SI}{Satisfaction Indicator}
  \acro{SIP}{Session Initiation Protocol}
  \acro{SISO}{single input single output}
  \acro{SIMO}{Single Input Multiple Output}
  \acro{SIR}{signal-to-interference ratio}
  \acro{SLNR}{Signal-to-Leakage-plus-Noise Ratio}
  \acro{SMA}{Simple Moving Average}
  \acro{SNR}{signal-to-noise ratio}
  \acro{SORA}{Satisfaction Oriented Resource Allocation}
  \acro{SORA-NRT}{Satisfaction-Oriented Resource Allocation for Non-Real Time Services}
  \acro{SORA-RT}{Satisfaction-Oriented Resource Allocation for Real Time Services}
  \acro{SPF}{Single-Carrier Proportional Fair}
  \acro{SRA}{Sequential Removal Algorithm}
  \acro{SRS}{Sounding Reference Signal}
  \acro{SSB}{synchronisation signal block}
  \acro{SU-MIMO}{single-user multiple input multiple output}
  \acro{SU}{Single-User}
  \acro{SVD}{Singular Value Decomposition}
  \acro{TCP}{Transmission Control Protocol}
  \acro{TDD}{time division duplexing}
  \acro{TDMA}{Time Division Multiple Access}
  \acro{TETRA}{Terrestrial Trunked Radio}
  \acro{TP}{Transmit Power}
  \acro{TPC}{Transmit Power Control}
  \acro{TTI}{Transmission Time Interval}
  \acro{TTR}{Time-To-Rendezvous}
  \acro{TSM}{Throughput-based Satisfaction Maximization}
  \acro{TU}{Typical Urban}
  \acro{Tx}{transmitter}
  \acro{UE}{user equipment}
  \acro{UEPS}{Urgency and Efficiency-based Packet Scheduling}
  \acro{UL}{uplink}
  \acro{ULA}{uniform linear array}
  \acro{UMTS}{Universal Mobile Telecommunications System}
  \acro{URI}{Uniform Resource Identifier}
  \acro{URM}{Unconstrained Rate Maximization}
  \acro{UT}{user terminal}
  \acro{VR}{Virtual Resource}
  \acro{VoIP}{Voice over IP}
  \acro{WAN}{Wireless Access Network}
  \acro{WCDMA}{Wideband Code Division Multiple Access}
  \acro{WF}{Water-filling}
  \acro{WiMAX}{Worldwide Interoperability for Microwave Access}
  \acro{WINNER}{Wireless World Initiative New Radio}
  \acro{WLAN}{Wireless Local Area Network}
  \acro{WMPF}{Weighted Multicarrier Proportional Fair}
  \acro{WPF}{Weighted Proportional Fair}
  \acro{WSN}{Wireless Sensor Network}
  \acro{WWW}{World Wide Web}
  \acro{XIXO}{(Single or Multiple) Input (Single or Multiple) Output}
  \acro{ZF}{zero-forcing}
  \acro{ZMCSCG}{Zero Mean Circularly Symmetric Complex Gaussian}
\end{acronym}

\begin{document}

\title{Rethinking Channel Charting: A Graph Perspective}

\author{\IEEEauthorblockN{Yifei Jin}
\IEEEauthorblockA{\textit{Ericsson Research}\\
Stockholm, Sweden \\
yifei.jin@ericsson.com}
\and
\IEEEauthorblockN{Yuxin Zhao}
\IEEEauthorblockA{\textit{Ericsson Research}\\
Linköping, Sweden \\
yuxin.zhao@ericsson.com}
\and
\IEEEauthorblockN{Dandan Hao}
\IEEEauthorblockA{\textit{Ericsson AB}\\
Stockholm, Sweden \\
dandan.hao@ericsson.com}
\and
\IEEEauthorblockN{G\'abor Fodor$^1$~\thanks{$^{1}$G. Fodor was supported by the Swedish Strategic Research (SSF) grant for the FUS21-0004 SAICOM project.}}
\IEEEauthorblockA{\textit{Ericsson Research}\\ \textit{KTH Royal Institute of Technology}\\
Stockholm, Sweden \\
gabor.fodor@ericsson.com~|~gaborf@kth.se}
}

\maketitle

\begin{abstract}
Channel charting is a self-supervised framework that learns low-dimensional spatial representations from high-dimensional channel state information. 
We revisit channel charting from a graph-theoretic perspective, and show that the position-diffusion objective is equivalent to a graph Laplacian smoothness functional. 
We propose a Graph Neural Network (GNN) formulation that replaces the Siamese network's global geodesic dissimilarity objective with a graph smoothness objective. We consider the reformulated objective to be the position diffusion objective. 
Without diverging from the original optimization objective, the GNN replaces the quadratic-cost self-correlation encoding with linear-cost message passing over an Angle-Delay Profile~(ADP)-similarity graph, achieving comparable positioning accuracy with $512\times$ fewer parameters. 
Using Laplacian spectral analysis, we demonstrate that obstacles compress the ADP graph spectrum, while the GNN acts as a spectral decompressor against obstacles and other environmental semantics, but as a compressor against excessive ADP embedding space. Beyond this, the second eigenvector of the learned embedding encodes the line-of-sight/non-line-of-sight boundary rather than spatial coordinates. 
\end{abstract}


\section{Introduction}
\label{sec:intro}

Recently, \ac{ML}-assisted positioning for 5G \ac{NR} systems has attracted considerable attention from both the research and standardization communities \cite{Butt:21, Ruan:23, Xue:24, Kang:25, Zha:25}. 
The \ac{3GPP}, for example, has developed \ac{RF} fingerprinting-based positioning approaches and control plane support for fingerprinting-assisted positioning~\cite{38843}. 
Although \ac{RF} fingerprinting is a promising approach for accurate localization in both indoor and outdoor scenarios, collecting sufficient amount of ground-truth (labeled) data in the training phase is costly. The amount of required labeled data limits the scalability across devices and renders practical applicability problematic~\cite{Wang:24, Zhao:24}.

Channel charting is a self-supervised technique that maps high-dimensional channel measurements to a low-dimensional chart, preserving the local spatial geometry of the radio environment~\cite{studer2018channelcharting}.
In contrast to \ac{RF} fingerprinting approaches, which typically rely on large labeled datasets and face scalability challenges as the number of devices and environments grows, channel charting learns spatial representations directly from unlabeled channel measurements through self-supervision~\cite{Tang:25, Jiao:26}.

Channel charts are structured by the geometry of the propagation environment: users that are close in physical space typically observe similar channel characteristics, whereas distant users experience increasingly distinct channels. Consequently, channel charting can be viewed as the problem of recovering a latent spatial manifold from local channel similarities. 
Such neighborhood relationships can be naturally represented as a graph, where nodes correspond to channel observations and edges encode local similarity.
This perspective suggests that graph-based learning methods may provide a more suitable inductive bias than architectures operating on individual channel samples \cite{Deng:21, Shaikh:25}.

To recover this latent geometry, channel charting methods rely on similarity measures derived directly from \ac{CSI}~\cite{Shaikh:24, Shaikh:25}. 
Among these, the \ac{ADP} dissimilarity has emerged as a particularly effective representation, as it captures propagation characteristics while exhibiting a strong correspondence with physical proximity under locally dominant \ac{LoS} or quasi-\ac{LoS} conditions \cite{Stephan:24}. 
By evaluating channel similarity across neighboring measurements, \ac{ADP} induces a weighted graph in which nodes correspond to \ac{CSI} observations and edges encode local spatial relationships. Existing embedding approaches~\cite{studer2018channelcharting,chaaya2024learning,taner2023channel} typically process channel measurements independently and recover spatial structure through pairwise objectives.

In contrast, \acp{GNN}~\cite{zhou2020graph} operate directly on the induced neighborhood graph, iteratively aggregating information from adjacent nodes through message passing. This inductive bias aligns naturally with the underlying assumptions of channel charting, where spatial information is encoded not only in individual channel measurements but also in their local relationships \cite{Shaikh:25, Thomas:24}.


In this paper, we revisit channel charting from a graph-theoretic perspective. We reformulate the \emph{global geodesic dissimilarity objective} formulated by~\citet{Shaikh:24} into a graph-centric \emph{position-diffusion objective}. We show that the \emph{position-diffusion objective} is equivalent to a \emph{graph Laplacian smoothness} functional whose gradient flow corresponds to a \emph{graph heat equation}.

Building on this interpretation, we propose a \ac{GNN}-based formulation that replaces explicit self-correlation encoding with message passing over an \ac{ADP}-similarity graph, reducing parameter complexity from quadratic to linear scaling with respect to the antenna dimensions. Finally, through Laplacian spectral analysis, we show that obstacles compress the \ac{ADP} graph spectrum while the proposed \ac{GNN} reveals the \ac{CSI}-related spatial features, which outperforms the baseline method. This perspective leads naturally to the formulation of channel charting as learning a low-dimensional embedding that preserves the local geometry encoded by channel similarities.

The remainder of this paper is organized as follows: Section~\ref{sec:background} reviews channel charting and \ac{ADP}-based similarity metrics. Section~\ref{sec:systemmodel} introduces the proposed graph formulation and \ac{GNN} architecture. Section~\ref{sec:experiments} presents experimental results and spectral analyses, while Section~\ref{sec:conclusion} concludes the paper.

\section{Background}
\label{sec:background}

\subsection{Channel Charting}
\label{subsec:ChannelCharting}

Channel charting~\cite{studer2018channelcharting} learns a low-dimensional spatial representation from unlabeled \ac{CSI} by exploiting the manifold assumption: in a static environment, \ac{CSI} is primarily a function of \ac{UE} position. The key enabler is a dissimilarity metric between measurements that correlates with physical distance. The \ac{ADP} metric~\cite{Stephan:24} computes per-tap cosine similarity across antennas and exhibits strong correspondence with Euclidean distance. State-of-the-art methods~\cite{Stephan:24, ferrand2020triplet, euchner2023augmenting,chaaya2024learning} embed \ac{CSI} via Siamese networks trained on geodesic \ac{ADP} distances over a $k$-\ac{NN} graph, processing each sample independently. However, jointly learning \ac{ADP} as a relational feature that associates each \ac{CSI} measurement has never been addressed in previous studies. \textbf{Given channel charting is a manifold, which can be discretized into a graph representation}, it is natural to revisit the channel charting in a graph perspective and study it through the \ac{GNN} method.

\subsection{Graph Neural Network}
\label{subsec:gnn}

\acp{GNN}~\cite{zhou2020graph, kipf2017semi,xupowerful} learn node representations through \emph{message passing}: iteratively aggregates features from graph neighbors, with trainable weights shared across all nodes. This inductive bias lies in the fact that \textbf{a node's representation depends on its local neighborhood}, which naturally suits the channel charting's smoothness objective~\cite{Stephan:24}, where spatial proximity implies \ac{CSI} similarity. Unlike Siamese networks that process samples independently, \acp{GNN} exploit both node features (i.e, \ac{CSI} measurement) and graph topology (i.e, \ac{ADP} or other inter-\ac{CSI} similarity metric) simultaneously. By accessing pairwise spatial information through edge structure rather than through explicit feature expansion, the proposed \ac{GNN} method achieved better performance with fewer parameters than the Siamese network.

\subsection{Graph Laplacian Analysis}
\label{subsec:graphlaplacian}
Graph Spectral Analysis~\cite{nica2018brief} has been a long-studied area in graph theory. One major contribution of Graph Spectral Analysis is the graph Laplacian, which builds the foundation of \ac{GCN}~\cite{kipf2017semi} and many other \acp{GNN}~\cite{xupowerful, zhou2020graph}, by giving a theoretical foundation on how the graph pattern evolves. Besides, the graph Laplacian also inspired graph cut, community detection, and a vast majority of the graph representation learning community. Among them, one research question is to leverage the graph Laplacian to reverse engineer the learnt embedding, to acquire a deep insight into what has been learnt through \ac{GNN} models~\cite{vasileiou2026position,hu2021graph}. Only~\citet{chaaya2024learning} denote that there exists a latent space for channel charting manifold, without further revealing what domain knowledge is encoded in such latent space. \textbf{In this paper, we found through graph Laplacian analysis that the proposed \ac{GNN} method is more informative on domain knowledge than the baseline method.}
\section{Problem Formulation and System Model}
\label{sec:systemmodel}
\subsection{Problem Formulation}
\label{subsec:problemformulation}
\xhdr{Problem formulation}
Formally, let $\mathcal{S} = \{(\mx{h}_i, \mx{p}_i)\}_{i=1}^{N}$ denote a collection of channel observations and their associated physical locations, where the positions are unavailable during training. Specifically, $\mathcal{S}$ denotes $N$ paired samples, where $\mx{h}_i \in \mathbb{C}^{B \times M \times N_{\mathrm{sub}}}$ is the \ac{CSI} ($B$ arrays, $M$ antennas, $N_{\mathrm{sub}}$ OFDM subcarriers) and $\mx{p}_i \in \mathbb{R}^{D}$ is the position (see Table~\ref{tab:notation}). The goal is to learn a projection $\mathcal{C}_\theta$, where $\theta$ is the parameter, such that:

\begin{table}[t]
\centering
\caption{Notation used in the paper.}
\label{tab:notation}
\footnotesize
\begin{tabular}{p{0.2\linewidth} p{0.72\linewidth}}
\hline
\textbf{Symbol} & \textbf{Description} \\
\hline
$\mx{h}_i \in \mathbb{C}^{B \times M \times N_{\mathrm{sub}}}$ & \ac{CSI} for sample $i$ \\
$B$, $M$, $N_{\mathrm{sub}}$ & Number of arrays, antennas per array, subcarriers \\
$b$, $m$, $\tau$ & Array index, antenna index, delay-tap index \\
$\mx{p}_i \in \mathbb{R}^{D}$ & Physical position of sample $i$ \\
$N$ & Dataset size \\
$\mathcal{S} = \{(\mx{h}_i, \mx{p}_i)\}$ & Dataset of paired samples \\
$\tilde{h}_{i,b,m,\tau}$ & Time-domain \ac{CIR} (via \ac{IDFT} of $\mx{h}_i$) \\
$\tau_{\min},\tau_{\max}$ & Informative tap range \\
$d_{\mathrm{ADP}}(\cdot,\cdot)$ & \ac{ADP} dissimilarity~\eqref{eq:adp} \\
$\mx{g}(\cdot)$, $\mx{f}(\cdot)$ & Feature extraction and embedding stages \\
$\mathcal{C}_\theta$ & Parameterized forward charting function \\
$G(V,E,X,Y)$ & $k$-NN graph: nodes $V$ (\ac{CSI}), edges $E$ (\ac{ADP}), attributes $X$, weights $Y$ \\
$\mathcal{N}(i)$ & Neighbourhood of node $v_i$ in graph $G$ \\
$\mathcal{P}_{ij}$ & Shortest path between $v_i$ and $v_j$ in $G$ \\
$a_{i,j}$ & Geodesic distance between nodes $v_i,v_j$ \\
$\tilde{\mx{A}}$ & Row-normalized adjacency ($\tilde{a}_{ij} = a_{ij}/\sum_k a_{ik}$) \\
$\tilde{\mx{L}}_0 = \mx{I} - \tilde{\mx{A}}$ & Normalized graph Laplacian \\
$\mx{P} \in \mathbb{R}^{N \times D}$ & Position matrix (stacked $\mx{p}_i$) \\
$\mx{X}^{(\ell)}$ & Node feature matrix at GNN layer $\ell$ \\
$\mx{W}^{(\ell)}$ & \ac{GNN} learnable weights at layer $\ell$ \\
$d_i$ & Degree of node $v_i$ in graph $G$ \\
$\mx{z}_i \in \mathbb{R}^d$ & Chart coordinates (embedding output) \\
$\mx{\mathfrak{v}}_i$, $\mathfrak{\lambda}_i$ & $i$th-Laplacian eigenvector/eigenvalue \\
\hline
\end{tabular}
\end{table}
\begin{equation}
\mathcal{C}_\theta:\; \mx{h}_i \;\longmapsto\; \mx{z}_i, 
\quad \text{s.t.}\quad |\mx{z}_i - \mx{z}_j|^2 \propto |\mx{p}_i - \mx{p}_j|^2.
\label{eq:cc_objective}
\end{equation}
\xhdr{Two-stage decomposition}
Most methods~\cite{studer2018channelcharting, Stephan:24, ferrand2020triplet,taner2023channel,euchner2023augmenting} decompose $\mathcal{C}$ into a feature extraction stage $\mx{g}(\cdot)$ and embedding stage $\mx{f}(\cdot)$:
\begin{equation}
\mx{h}_i \xmapsto{\;\mx{g}(\cdot)\;} \mx{x}_i \in \mathbb{R}^{n}
    \xmapsto{\;\mx{f}(\cdot)\;} \mx{z}_i \in \mathbb{R}^{d},
\quad d \ll n \ll \dim(\mx{h}_i).
\label{eq:two_stage}
\end{equation}
Note by \citet{studer2018channelcharting}, $\mx{f}(\cdot)$ should be \emph{convex} and \emph{smooth}.
The design of $\mx{g}(\cdot)$ determines what feature representation space is available to $\mx{f}(\cdot)$: geometric parameters~\cite{euchner2023augmenting}, pairwise dissimilarities~\cite{Stephan:24,ferrand2020triplet}, or graph-aggregated features (this paper). 

\subsection{System Model}
\label{subsec:systemmodel}
\xhdr{Channel Charting via \ac{ADP} Dissimilarity}
\label{subsubsec:ADP}
In the feature extraction stage $\mx{g}(\cdot)$, \ac{ADP} has been one of the major feature spaces to leverage \ac{CSI} measurement (dis-)similarity.
Applying the \ac{IDFT} to \ac{CSI} yields the \ac{CIR} $\tilde{h}_{i,b,m,\tau}$, where $b \in \{1,\ldots,B\}$ indexes antenna arrays, $m \in \{1,\ldots,M\}$ indexes antennas within each array, and $\tau \in [\tau_{\min}, \tau_{\max}]$ is the delay-tap index covering the dominant \acp{MPC}. The \ac{ADP} dissimilarity~\cite{Stephan:24} computes per-tap cosine similarity across antennas:
\begin{equation}
d_{\mathrm{ADP}}(\mx{h}_i, \mx{h}_j) \triangleq 
\sum_{b,\tau} \left(1 - \frac{|\sum_{m} \tilde{h}_{i,b,m,\tau}^{*} \tilde{h}_{j,b,m,\tau}|^{2}}{\sum_{m}|\tilde{h}_{i,b,m,\tau}|^{2} \cdot \sum_{m}|\tilde{h}_{j,b,m,\tau}|^{2}}\right).
\label{eq:adp}
\end{equation}

A $k$-\ac{NN} graph built from $d_{\mathrm{ADP}}$ with geodesic shortest-path distances approximates Euclidean distances when a \ac{LoS} or strong quasi-\ac{LoS} component dominates locally~\cite{euchner2023augmenting}.

\xhdr{Siamese Network Embedding}
\label{subsubsec:SiameseNN}
The state-of-the-art embedding $\mx{f}(\cdot)$ for channel charting uses a Siamese network~\cite{Stephan:24,ferrand2020triplet}: two weight-sharing branches process \ac{CSI} pairs $(\mx{h}_i, \mx{h}_j)$, trained to match predicted embedding distance to geodesic \ac{ADP} distance. Each branch applies a \emph{FeatureEngineering} layer that computes the full antenna sample covariance:
\begin{equation}
\mathrm{ac}[t, a, b, m, n] = \tilde{h}_i[a, m, t] \cdot \tilde{h}_i^*[b, n, t]; a,b \in [B], m,n\in [M],
\label{eq:outer_product}
\end{equation}
producing a tensor of shape $(\tau_{\max}\!-\!\tau_{\min}) \times B^2 \times M^2 \times 2$ (real/imaginary). This is flattened and compressed via dense layers to chart coordinates $\mx{z}_i \in \mathbb{R}^2$. The outer product~\eqref{eq:outer_product} is necessary because the Siamese processes each sample \emph{independently} as it must extract all angular/spatial information from a single measurement.

\xhdr{Graph Neural Network}
\label{subsubsec:gnn}
In contrast to previous papers, we introduce a novel implementation of embedding $\mx{f}(\cdot)$ through \ac{GNN}. \acp{GNN} learn node representations by iteratively aggregating features from graph neighbors, which is the same \ac{ADP} matrix, but just considered as an adjacency matrix connecting nodes that represent different locations' \ac{CSI} measurements.

Formally, we reformulate the $\mathcal{S} = \{(\mx{h}_i, \mx{p}_i)\}_{i=1}^{N}$ as a weighted, attributed graph $G(V,E,X,Y)$. Each node $v_i\in V$ denotes a \ac{CSI} measurement, with node attribute $\mx{x}(v_i)\rightarrow \mx{x}_i \in X$ denotes its value $\mx{h}_i$. Each edge $e_{i,j} \in E$ denotes a \ac{ADP} dissimilarity relation $d_{\mathrm{ADP}}(\mx{h}_i, \mx{h}_j)$, with edge weight $y(e_{i,j})\rightarrow a_{i,j}\in Y$ denotes its value. The learning objective is to learn an embedding for each node, given an $N$-node graph $G(V, E, X, Y)$. Naturally, we consider a \ac{GCN} architecture to replace the Siamese architecture, a \ac{GCN} layer updates node $v_i$'s embedding $\mx{x}_i^{(\ell)}\rightarrow \mx{x}_i^{(\ell+1)}$ as:
\begin{equation}
\mx{x}_i^{(\ell+1)} = \sigma\!\left(\mx{W}^{(\ell)}\mx{x}_i^{(\ell)} + \mx{W}^{(\ell)}\sum_{j \in \mathcal{N}(i)} \frac{a_{i,j}}{\sqrt{d_i d_j}}\,  \mx{x}_j^{(\ell)}\right),
\label{eq:gnn_layer}
\end{equation}
where $d_i, d_j$ are the degree of node $v_i, v_j$ and trainable weight $\mx{W}^{(\ell)}$ is shared across all nodes. Eq.~\ref{eq:gnn_layer} brings an inductive bias: \textbf{a node's representation depends on its local neighborhood, is naturally suited to channel charting} (i.e, spatial proximity implies \ac{CSI} similarity).

\section{Methodology: \ac{GNN}-Based Channel Charting}
\label{sec:method}

\subsection{Graph Construction}
\label{subsubsec:graphconstruction}
Given the graph $G(V,E,X,Y)$ defined in Section~\ref{subsubsec:gnn}, we sparsify $E$ by retaining only the $k$=20 nearest neighbors per node in $d_{\mathrm{ADP}}$ space. The \ac{GCN} message passing in equation ~\eqref{eq:gnn_layer} operates on these \emph{local} \ac{ADP} weights. Similar to the geodesic distance definition in~\cite{Stephan:24}, we rewrite it in a graph form:
\begin{equation}
a_{i,j} = \min_{\mathcal{P}_{ij}} \sum_{e_{m,n} \in \mathcal{P}_{ij}} d_{\mathrm{ADP}}(\mx{h}_m, \mx{h}_n),
\label{eq:geodesic}
\end{equation}
which chains local \ac{ADP} along shortest paths $\mathcal{P}_{ij}$ between $v_i$ and $v_j$ to approximate Euclidean distance globally. Note that $a_{i,j}$ in~\eqref{eq:geodesic} gives a scalar distance for each node pair $(v_i, v_j)$.
Eq.~\ref{eq:geodesic} relies on the following Lemma~\ref{lem:local}, by~\citet{studer2018channelcharting} and~\citet{Stephan:24}, but reformulated in differential geometry form:
\begin{lemma}[Local distance-dissimilarity correspondence]
\label{lem:local}
If the dissimilarity function $d_{\mathrm{ADP}}(\cdot,\cdot)$ is smooth in physical distance~\cite{studer2018channelcharting}, and the graph $G$ is constructed with sufficiently dense sampling such that connected nodes $e_{m,n} \in E$ satisfy $\|\mx{p}_m - \mx{p}_n\|_2 = \varepsilon \to 0$, then on a Riemannian manifold with sectional curvature $\kappa$~\cite{ollivier2009ricci}:
\begin{equation}
a_{m,n} = c\|\mx{p}_m - \mx{p}_n\|_2 + \mathcal{O}(\kappa\varepsilon^3),
\label{eq:taylor_local}
\end{equation}
i.e., the geodesic hop approximates Euclidean distance up to a curvature-dependent residual that vanishes for locally flat manifolds ($\kappa \approx 0$) or dense sampling ($\varepsilon \to 0$).
\end{lemma}

For the \emph{training objective}, derived from the geodesic distance formulation in Eq.~\ref{eq:geodesic}, we also rewrite in graph form:
\begin{equation}
\label{eq:geodesicobjective}
\min_{\theta} \sum_{v_i, v_j\in V} (a_{i,j} - \|\mathcal{C}_\theta(\mx{h}_i) - \mathcal{C}_\theta(\mx{h}_j)\|_2)^2
\end{equation}

\subsection{Position Diffusion as Graph Heat Equation}
\label{subsec:positiondiffusion}
\xhdr{Global to Local Smoothness} Different from previous work that optimizes a geodesic dissimilarity objective as per Eq.~\ref{eq:geodesicobjective}, we reformulate this into a \emph{position diffusion objective}. Substituting Eq.~\ref{eq:cc_objective} and~\ref{eq:geodesic} into Eq.~\ref{eq:geodesicobjective}, the objective becomes:
\begin{equation}
\label{eq:derivedgeodesicobjective}
\min_\theta \sum_{v_i, v_j\in V} [(\min_{\mathcal{P}_{ij}} \sum_{e_{m,n} \in \mathcal{P}_{ij}} a_{m,n}) - c\|\mx{p}_i - \mx{p}_j\|_2]^2
\end{equation}
where $c$ is a constant. By Lemma~\ref{lem:local}, each hop satisfies $a_{m,n} = c\|\mx{p}_m - \mx{p}_n\|_2 + \mathcal{O}(\kappa\varepsilon^3)$. For a path of $L$ hops on a locally flat manifold ($\kappa \approx 0$):
\begin{equation}
\sum_{e_{m,n} \in \mathcal{P}_{ij}} a_{m,n} = c\sum_{e_{m,n} \in \mathcal{P}_{ij}} \|\mx{p}_m - \mx{p}_n\|_2 + L\cdot\mathcal{O}(\kappa\varepsilon^3).
\label{eq:path_taylor}
\end{equation}
On a flat manifold, when the shortest graph path follows the Euclidean geodesic (straight line), the sum of hop lengths equals the total displacement: $\sum \|\mx{p}_m - \mx{p}_n\|_2 = \|\mx{p}_i - \mx{p}_j\|_2$. Thus, Eq.~\ref{eq:derivedgeodesicobjective} is satisfied (up to $\mathcal{O}(L\kappa\varepsilon^3)$) whenever positions are \emph{harmonic}, expressed as:
\begin{equation}
\tilde{\mx{L}}_0\mx{P} = 0 \quad \Longleftrightarrow \quad \mx{p}_i = \sum_{j \in \mathcal{N}(i)} \tilde{a}_{ij}\,\mx{p}_j, \quad \forall\, v_i \in V,
\label{eq:harmonic}
\end{equation}
where $\tilde{\mx{L}}_0 := \mx{I} - \tilde{\mx{A}}$ denote as the normalized graph Laplacian, $\tilde{\mx{A}}$ is the row-normalized adjacency ($\tilde{a}_{ij} = a_{ij}/\sum_k a_{ik}$). Stacking p//ositions into $\mx{P} \in \mathbb{R}^{N \times D}$, deviation from harmonicity is:
\begin{equation}
\min_{\mx{P}}\; \|\tilde{\mx{A}}\mx{P} - \mx{P}\|_F^2,
\label{eq:diffusionobjective}
\end{equation}
which we define as the \textbf{position diffusion objective}.

\xhdr{Smoothness to Graph Heat Equation} The position diffusion objective~\eqref{eq:diffusionobjective} has an equivalent spectral form:
\begin{equation}
\min_{\mx{P}}\|\tilde{\mx{A}}\mx{P} - \mx{P}\|_F^2 
= \min_{\mx{P}}\mathrm{tr}(\mx{P}^T \tilde{\mx{L}}_0^2 \mx{P})
\xrightarrow{\tilde{\mx{L}}_0 \succeq 0}
\min_{\mx{P}}\; \mathrm{tr}(\mx{P}^T \tilde{\mx{L}}_0 \mx{P}),
\label{eq:O2}
\end{equation}
where the shared minimizer is $\tilde{\mx{L}}_0\mx{P} = 0$ (since $\tilde{\mx{L}}_0$ is positive semi-definite). The first-order form has gradient flow equal to the \textbf{graph heat equation}~\cite{chen2012heat}:
\begin{equation}
-\nabla_{\mathbf{P}} \mathrm{tr}\!\left(\mathbf{P}^T \tilde{\mx{L}}_0 \mathbf{P}\right) = -\tilde{\mx{L}}_0\,\mx{P}^{(\ell)} = \mx{P}^{(\ell+1)} - \mx{P}^{(\ell)}.
\label{eq:heat_eq}
\end{equation}
The \ac{GNN} (Eq.~\ref{eq:gnn_layer}) matrix form $\mx{X}^{(\ell+1)} = \sigma(\tilde{\mx{A}}\mx{X}^{(\ell)}\mx{W}^{(\ell)})$ reveals that each \ac{GNN} layer is a \emph{learnable one-step diffusion}: the heat equation propagates with fixed dynamics, while the \ac{GNN} applies parametric diffusion through $\mx{W}^{(\ell)}$ at each step.

\subsection{\ac{GNN} vs.\ Siamese: Architectural Comparison}
\label{subsec:comparison}
Both methods map \ac{CSI} to 2D chart coordinates but differ fundamentally in how spatial information is extracted:

\noindent\textbf{Siamese (self-correlation):}
\begin{equation}
\mx{h}_i \in \mathbb{C}^{B \times M \times \tau}
\xrightarrow{\mx{h}\mx{h}^*}
\mathbb{R}^{2B^2 M^2 \tau}
\xrightarrow{\mx{W}_1}
\mathbb{R}^{512}
\xrightarrow{\mathrm{Dense}}
\mx{z}_i \in \mathbb{R}^{2}
\label{eq:Siamese_flow}
\end{equation}

\noindent\textbf{GNN (message passing):}
\begin{equation}
\mx{h}_i \in \mathbb{C}^{B \times M \times \tau}
\xrightarrow{\mathrm{Flatten}}
\mathbb{R}^{2BM\tau}
\xrightarrow{\mx{W}_1}
\mathbb{R}^{512}
\xrightarrow{\mathrm{GCN} \times L}
\mx{z}_i \in \mathbb{R}^{2}
\label{eq:gnn_flow}
\end{equation}

\begin{figure}[t]
\centering
\includegraphics[width=0.9\columnwidth]{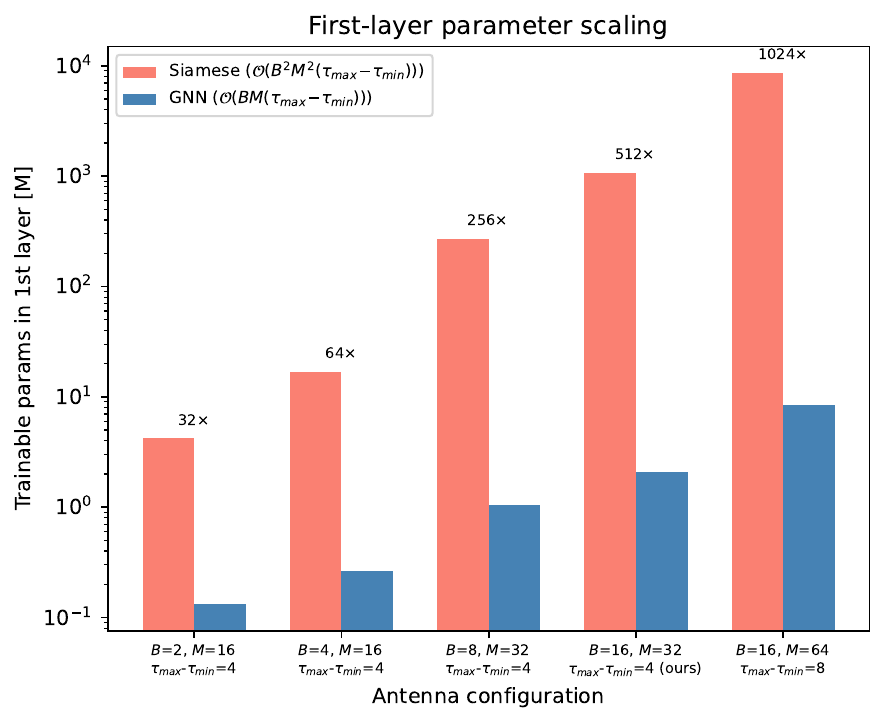}
\caption{First-layer parameter scaling. The Siamese outer product causes $\mathcal{O}(B^2 M^2)$ growth; the \ac{GNN} scales linearly as $\mathcal{O}(BM)$.}
\label{fig:param_scaling}
\end{figure}

The bottleneck layer $\mx{W}_1$ scales as $\mathcal{O}(B^2 M^2 \tau \cdot d)$ for the Siamese versus $\mathcal{O}(BM\tau \cdot d)$ for the \ac{GNN} (Fig.~\ref{fig:param_scaling}). The \ac{GNN} avoids the outer product because graph message passing implicitly encodes pairwise spatial information. Referring Eq.~\eqref{eq:gnn_layer}, the first term is a learnt projection of the node's own \ac{CSI} (subsuming self-correlation), while the second is a learnt \emph{cross-correlation} with the neighborhood-averaged \ac{CSI}. The \ac{ADP} edge weights ensure this cross-correlation is spatially meaningful.

\section{Experiments and Spectral Analysis}
\label{sec:experiments}

\subsection{Setup}
\label{subsec:Setup}
We evaluate on ray-traced indoor scenes using \textsc{NVIDIA Sionna}~\cite{hoydis2022sionna}: one empty room, one $1$m metallic cube at room centre, and a $3$m metallic wall at mid-position towards room centre. Each scene has $25600$ UE positions on an $8\times8$m grid, with a single base station. For each scene, we consider pure specular reflection or with diffraction ray setup. For neural architecture: \ac{GNN}: $3$-layer \textsc{GCNConv}~\cite{kipf2017semi}, $k$=30. Siamese: \textsc{FeatureEngineering} layer + \textsc{Dense} layer, with neuron size $[512,256,128]$. Both use $80/20$ train/test split.

\subsection{Positioning Accuracy}
\label{subsec:accuracy}
\begin{table}[t]
\centering
\caption{Positioning accuracy (\ac{MAE} [m]).}
\label{tab:positioning}
\begin{tabular}{lccc}
\hline
Environment & Siamese & \ac{GNN} & Params. \\
\hline
Empty room spec. & 0.562 & \textbf{0.541} & \multirow{6}{*}{\shortstack{2.45M (GNN)\\vs.\ $\sim$1B (Siam.)}} \\
Empty room diff. & 0.570 & \textbf{0.538} & \\
Cube center spec. & 0.847 & \textbf{0.833} & \\
Cube center diff. & \textbf{0.811} & 0.813 & \\
Wall mid spec. & 0.607 & \textbf{0.582} & \\
Wall mid diff. & 0.608 & \textbf{0.560} & \\
\hline
\end{tabular}
\end{table}

\begin{figure*}[t]
\centering
\begin{subfigure}[t]{0.8\textwidth}
\centering
\includegraphics[width=\textwidth]{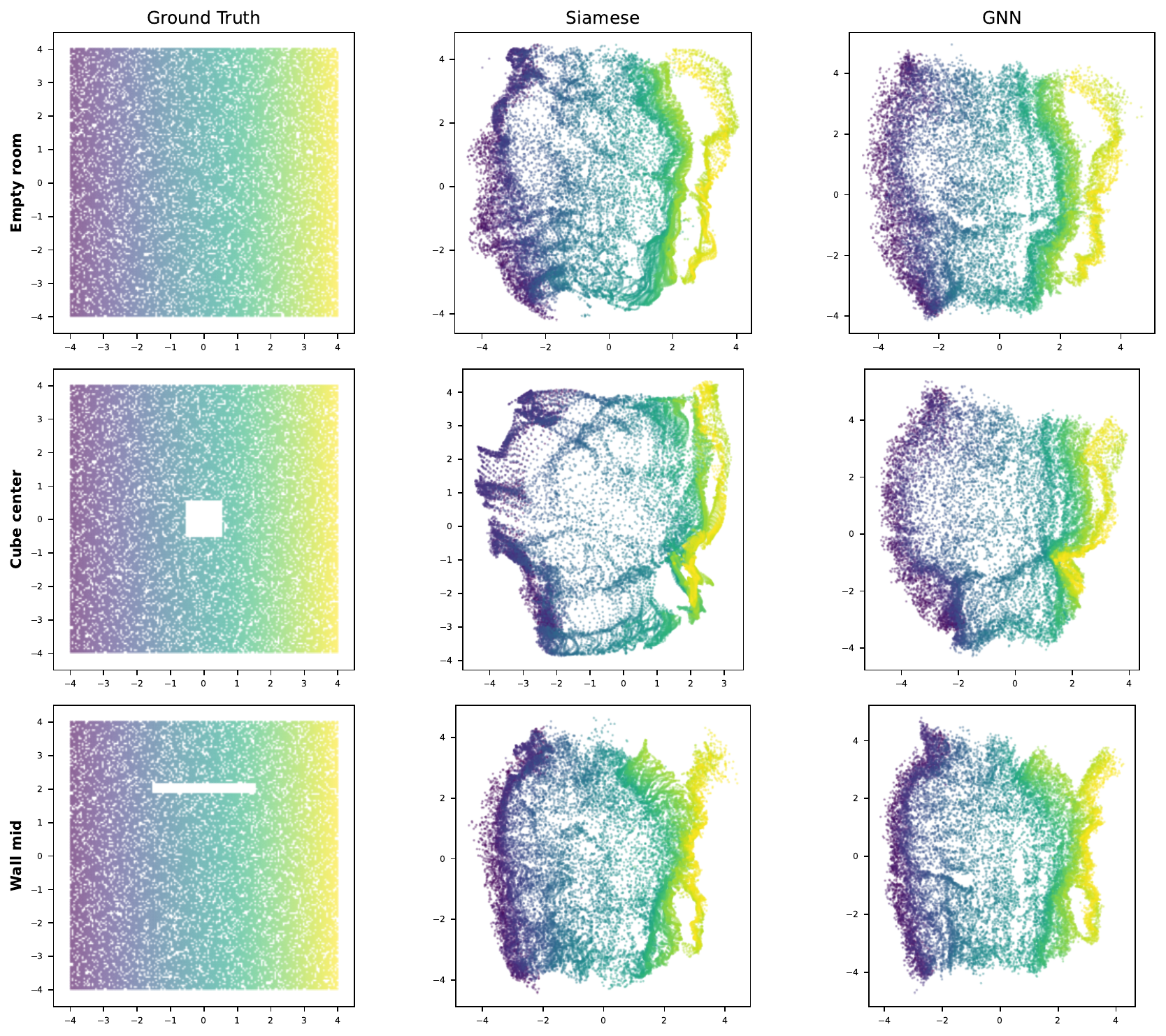}
\caption{Channel charts projected to physical coordinates, colored by x-position. Rows: Empty room, Cube center, Wall mid. Columns: ground truth, Siamese, \ac{GNN}.}
\label{fig:charts_comparison}
\end{subfigure}
\vspace{0.3em}
\begin{subfigure}[t]{0.8\textwidth}
\centering
\includegraphics[width=\textwidth]{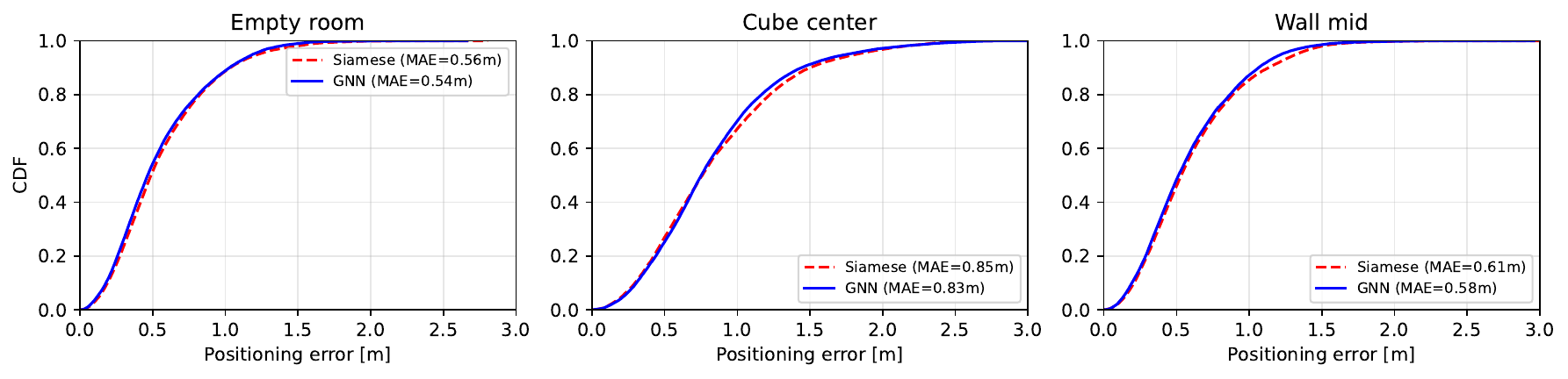}
\caption{Error \ac{CDF} per environment. Siamese (dashed red) vs.\ \ac{GNN} (solid blue). \ac{MAE} values in legend.}
\label{fig:error_cdf}
\end{subfigure}
\caption{Positioning results (Specular reflection only; diffraction is similar.). (a) Learnt channel charts. (b) Error \ac{CDF}.}
\label{fig:positioning_results}
\end{figure*}

Table~\ref{tab:positioning} shows the \ac{GNN} matches or outperforms the Siamese across all environments with $512\times$ fewer parameters. 
The improvement is largest in obstacle scenes, where the graph structure provides context that independent sample processing cannot. Fig.~\ref{fig:charts_comparison} visualizes the learnt charts, and Fig.~\ref{fig:error_cdf} shows the per-environment error \ac{CDF}.

\subsection{Spectral Analysis}

Given that we learn an updated charting space from $G(V,E,X,Y)$, we applied Laplacian analysis to each learnt embedding graph, to compare the learning outcome against the raw \ac{ADP} input formed graph. We construct four $k$-NN graphs ($k$=30) on the same nodes: $G_{\mathrm{GT}}$ (positions), $G_{\mathrm{ADP}}$ (ADP), $G_{\mathrm{Siam}}$ (Siamese chart), $G_{\mathrm{GNN}}$ (GNN chart), with edge weight $1/|\mx{z_i} - \mx{z_j}|^2$ or $d_{\mathrm{ADP}}(\mx{h}_i, \mx{h}_j)$ (only for $G_{\mathrm{ADP}}$), and compute normalized Laplacian spectra.

\begin{figure*}[t]
\centering
\begin{subfigure}[t]{\textwidth}
\centering
\includegraphics[width=0.8\textwidth]{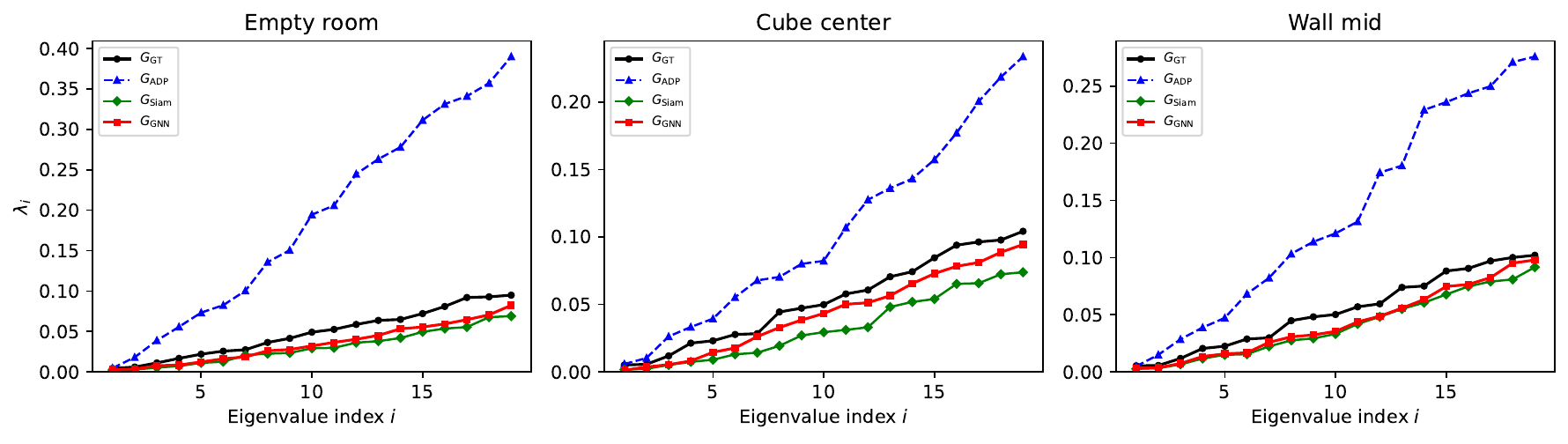}
\caption{Laplacian spectrum across three environments: $G_{\mathrm{GT}}$ (black), $G_{\mathrm{ADP}}$ (blue), $G_{\mathrm{Siam}}$ (green), $G_{\mathrm{GNN}}$ (red).}
\label{fig:laplacian_spectrum}
\end{subfigure}
\vspace{0.2em}
\begin{subfigure}[t]{\textwidth}
\centering
\includegraphics[width=0.8\textwidth]{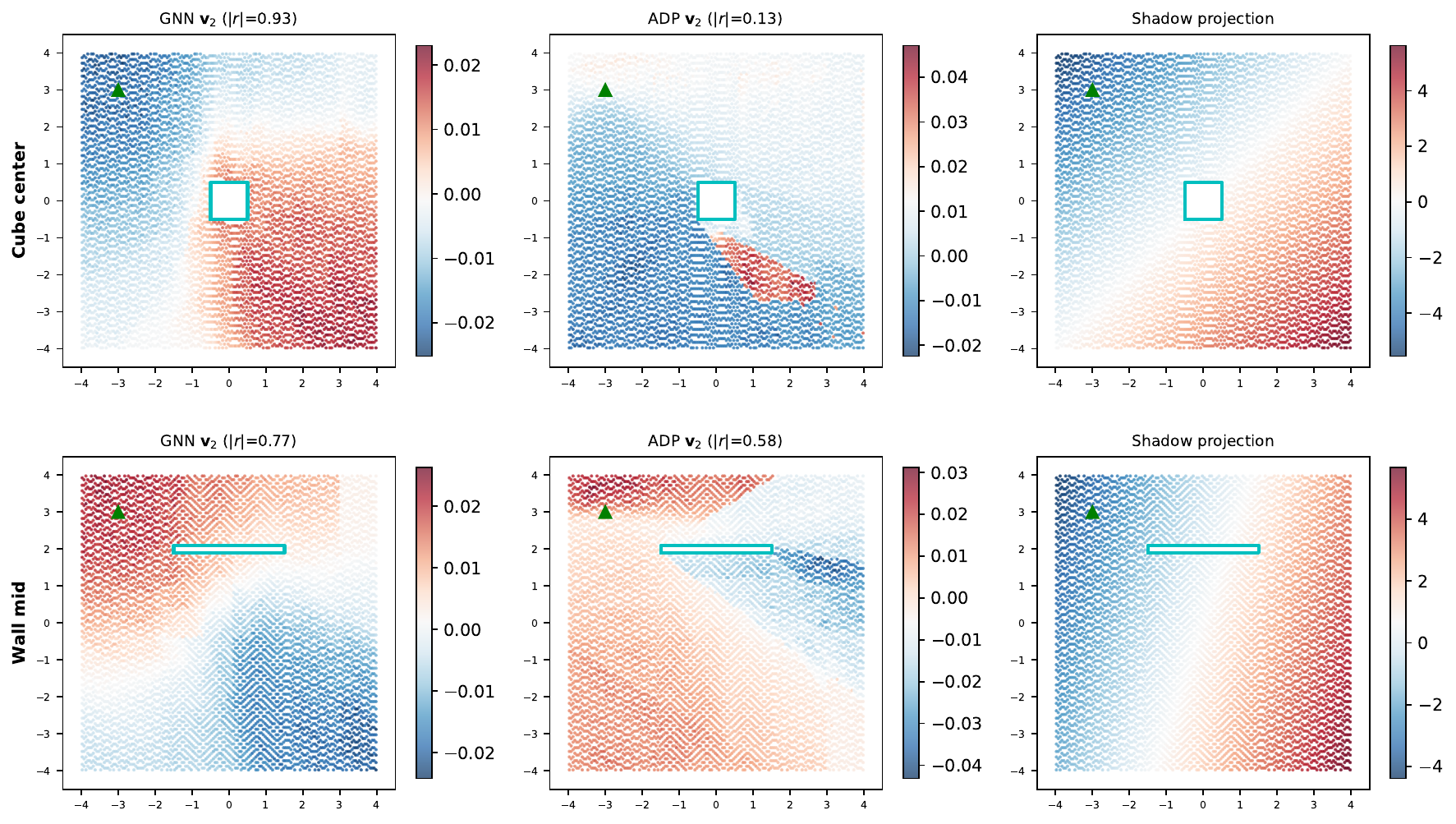}
\caption{Second eigenvector: $G_{\mathrm{GNN}}$ $\mx{\mathfrak{v}}_2$ vs $G_{\mathrm{ADP}}$ $\mx{\mathfrak{v}}_2$ vs shadow projection (signed distance along TX$\to$obstacle axis).}
\label{fig:v2_shadow}
\end{subfigure}
\caption{Spectral analysis. (a) \ac{GNN} decompresses the \ac{ADP} spectrum toward GT. (b) \ac{GNN} $\mx{\mathfrak{v}}_2$ encodes the \ac{LoS}/N\ac{LoS} boundary.(Only the spectral reflection plotting, with the diffraction plotting is similar.)}
\label{fig:spectral_analysis}
\end{figure*}

Fig.~\ref{fig:spectral_analysis}(a): Within each environment, $G_{\mathrm{ADP}}$ exhibits a larger-valued Laplacian spectrum than $G_{\mathrm{GNN}}$, which in turn is closer to $G_{\mathrm{GT}}$. This indicates the \ac{GNN} acts as a \emph{spectral compressor} that maps the inflated \ac{ADP} dissimilarity space toward the ground-truth spatial geometry. Across environments, obstacles compress the $G_{\mathrm{ADP}}$ mid-spectrum relative to the empty room; the \ac{GNN} \emph{decompresses} these obstacle-affected eigenvalues back toward $G_{\mathrm{GT}}$, restoring spatial information corrupted by introducing the obstruction.

Fig.~\ref{fig:spectral_analysis}(b): A notable \emph{eigengap} between $\mathfrak{\lambda}_2$ and $\mathfrak{\lambda}_3$ in Fig.~\ref{fig:laplacian_spectrum} suggests that $\mx{\mathfrak{v}}_2$ captures the dominant non-trivial structure of the embedding graph. The second eigenvector $\mx{\mathfrak{v}}_2$ of $G_{\mathrm{GNN}}$ correlates with the obstacle's shadow direction ($|r|>0.77$, Table~\ref{tab:v2_alignment}), while $G_{\mathrm{ADP}}$'s $\mx{\mathfrak{v}}_2$ does not and $G_{\mathrm{Siam}}$'s is not consistent or less informative. The \ac{GNN} devotes its leading non-trivial axis to encoding the \ac{LoS}/N\ac{LoS} boundary.

\begin{table}[t]
\centering
\caption{$|r|$ between $\mx{\mathfrak{v}}_2$ and spatial features ($N$=7000).}
\label{tab:v2_alignment}
\footnotesize
\begin{tabular}{llccc}
\hline
Scenario & Feature & GNN $\mx{\mathfrak{v}}_2$ & Siam.\ $\mx{\mathfrak{v}}_2$ & ADP $\mx{\mathfrak{v}}_2$ \\
\hline
\multirow{2}{*}{Cube center spec.} & Shadow & \textbf{0.93} & 0.14 & 0.13 \\
 & AoA & \textbf{0.71} & 0.53 & 0.30 \\
\hline
\multirow{2}{*}{Cube center diff.} & Shadow & \textbf{0.92} & 0.13 & 0.31 \\
 & AoA & \textbf{0.47} & 0.34 & 0.25 \\
\hline
\multirow{2}{*}{Wall mid spec.} & Shadow & 0.77 & \textbf{0.92} & 0.58 \\
 & AoA & 0.56 & \textbf{0.67} & 0.32 \\
\hline
\multirow{2}{*}{Wall mid diff.} & Shadow & \textbf{0.80} & 0.74 & 0.52 \\
 & AoA & \textbf{0.53} & 0.50 & 0.41 \\
\hline
\end{tabular}
\end{table}

\section{Conclusion}
\label{sec:conclusion}

We revisited channel charting from a graph perspective, establishing that position diffusion is a Laplacian smoothness functional whose gradient flow equals the graph heat equation. The \ac{GNN} replaces the Siamese's $\mathcal{O}(B^2M^2)$ self-correlation with $\mathcal{O}(BM)$ message passing, achieving a $512\times$ parameter reduction. Spectral analysis reveals the \ac{GNN} acts as a spectral decompressor and that obstacle-induced boundaries dominate the learnt embedding structure. More generally, our work suggests that channel charting should be viewed as graph representation learning rather than metric learning. 

\printbibliography
\medskip


\end{document}